\documentclass[twocolumn,spconf,aps,prl,a4paper,
  floatfix,
  groupedaddress,
  superscriptaddress,
  preprint,preprintnumbers,
  showpacs,showkeys,
  rmp,amsmath,amssymb]{revtex4-1}

\usepackage{physics}
\usepackage{setspace}
\usepackage{calc}
\usepackage{graphicx}
\usepackage{color}
\usepackage[dvips]{epsfig}  
\usepackage{bm}
\usepackage{bbold}
\usepackage[T1]{fontenc}
\usepackage{txfonts}
\usepackage{chngcntr} 
\usepackage{hyperref}
\usepackage{algorithm}
\usepackage{algorithmic}
\usepackage{subcaption}
\usepackage{cleveref}
\usepackage{dsfont}
\usepackage{relsize}
\usepackage[titletoc]{appendix}

\newcommand{\TIFR}{Department of Biological Sciences, Tata Institute of Fundamental Research, Mumbai, Maharashtra 400005, India}
\newcommand{\CBS}{UM-DAE Centre for Excellence in Basic Sciences, Biological Sciences, Kalina campus, Santacruz (E), Mumbai,  Maharashtra 400098, India}

\counterwithout{equation}{section}
\counterwithout{equation}{subsection}
\counterwithout{equation}{subsubsection}

\begin{document}


\title{Unified theory of human genome reveals a constrained spatial chromosomal arrangement in interphase nuclei}


\author{Sarosh N. Fatakia}
\altaffiliation{Co-corresponding author (SNF) email}
\email[]{sarosh.fatakia @ tifr.res.in}
\affiliation{\TIFR}

\author{Ishita S. Mehta}
\affiliation{\TIFR}
\affiliation{\CBS}

\author{Basuthkar J. Rao}
\altaffiliation{Co-corresponding author (BJR) email}
\email[]{bjrao@ tifr.res.in}
\affiliation{\TIFR}



\begin{abstract}
We investigate a densely packed, non-random arrangement of forty-six chromosomes ($46$,XY) in human nuclei. Here, we model systems-level chromosomal crosstalk by unifying intrinsic parameters (chromosomal length and number of genes) across all pairs of chromosomes in the genome to derive an extrinsic parameter called effective gene density. The hierarchical clustering and underlying degeneracy in the effective gene density space reveal systems-level constraints for spatial arrangement of clusters of chromosomes that were previously unknown. Our findings corroborate experimental data on spatial chromosomal arrangement in human nuclei, from fibroblast and lymphocyte cell lines, thereby establishing that human genome constrains chromosomal arrangement. We propose that this unified theory, which requires no additional experimental input, may be extended to other eukaryotic species, with annotated genomes, to infer their constrained self-organized spatial arrangement of chromosomes.
\end{abstract}

\keywords{chromosome crosstalk; chromosome arrangement; chromosome territory; effective gene density; human genome; radial chromosomal arrangement; self-organized chromosomal arrangement; systems biology; systems biology constraints for self-organized CT arrangement; systems-level chromosome crosstalk}

\maketitle

\def\aver#1{{\left<{#1}\right>}}
\def\Rset{\hbox{{I\kern -0.2em R}}}
\def\rset{\hbox{{\tiny\rm I\kern -0.2em R}}}
\def\udm{{1\over2}}
\def\emt{{e^{-\alpha\t}}}
\def\emdt{{e^{-2\alpha\t}}}
\def\expo#1{{e^{^{{#1}}}}}
\def\kb{{k_{\hbox{\tiny B}}}}
\def\Te{{T_e}}
\def\kT{{\kb\Te}}
\def\capa{{\mathcal C_{\rho_H}}}
\def\capax{{\mathcal C_{x}}}
\def\CTjk{{C_{jk}^{'}}}
\def\CTkj{{C_{kj}^{'}}}
\def\CTj{C_{j}}
\def\CTk{C_{k}}
\def\Dmatrix{\mathit {\hat{D}}}

%

\renewcommand{\epsilon}{\varepsilon}
\renewcommand{\leq}{\leqslant}
\renewcommand{\geq}{\geqslant}
\newcommand{\myPlus}{{\boldsymbol{+}}}
\newcommand{\om}{{\boldsymbol{\omega}}}
\newcommand{\bSigma}{{\boldsymbol{\Sigma}}}
\newcommand{\bTau}{{\boldsymbol{\Tau}}}
\newcommand{\bGamma}{{\boldsymbol{\Gamma}}}
\newcommand{\bp}{{\boldsymbol{p}}}
\newcommand{\bP}{{\boldsymbol{P}}}
\newcommand{\bD}{{\boldsymbol{D}}}
\newcommand{\bL}{{\boldsymbol{L}}}
\newcommand{\bbrackl}{{\[\left[}}
\newcommand{\bbrackr}{{\[\right]}}
\newcommand{\aUnit}{1\!\!1} 

\newcommand{\matr}[1]{\mathbf{#1}} 
\newcommand{\bbrackll}{{\left[}}
\newcommand{\bbrackrr}{{\right]}}
\def\r{{\mathbf{r}}}
\def\R{\mathbf{R}}
\renewcommand{\Rset}{\mathbb{R}}


\begin{section}{Introduction}
The human nucleus is a densely packed many-body system with forty-six chromosomes, each of which are manifest as individual physical units called chromosome territories (CTs) \cite{Cremer_2010}, during interphase stage of its cell cycle. Qualitative and quantitative microscopy data have revealed a characteristic, radially arranged pattern (reviewed in \cite{Cremer_2010,Misteli_2010,Bickmore_2013_Ann_Rev}) of CTs in human nuclei (\cite{Bolzer_2005,Boyle_2001,Cremer_2001,Croft_1999,Mehta_2013,Sun_2000}, and corroborated at our laboratory using high volume data \cite{Mehta_Schak_2013}), and in other species, particularly higher eukaryotes (reviewed in \cite{Foster_2005}). CTs are spatially and temporally regulated entities because experimental evidence has demonstrated that normal chromosomal function significantly impacts its position and vice versa (reviewed in \cite{Bickmore_2013_Cell, Pombo_2015, Sexton_2015}). For example, it has been recently established in our laboratory, using human dermal fibroblast cells, that some CTs undergo statistically significant displacement (relocate from their original to a new position in the nucleus) after controlled induction of DNA double strand breaks by cisplatin (DNA damaging agent) treatment \cite{Mehta_2013}. A significant population of cells showed that chr17 and chr19 relocated from the interior of the nucleus to the periphery, while chr12 and chr15 were displaced in an opposite sense (from the periphery toward the interior). At the population-level, statistically insignificant displacement was also recorded for chr20 that is largely located at the interior. Original CT arrangement was restored upon washing off cisplatin from the treated cells following DNA repair \cite{Mehta_2013}. Independent studies have reported that DNA double strand breaks lead to instances of chromosomal translocations specifically among closely located chromosomes in cancers and other genetic instability disorders (\cite{Roix_2013} and reviewed in \cite{Roukos_2014}). Therefore, spatial non-random arrangement of chromosomes within the nucleus directly influences not only the chromosomal functions (genetic and epigenetic) but also their translocation propensities \cite{Roukos_2014}. As CT arrangement significantly differs in tumor versus normal cells \cite{Parada_2002}, it is critical to unravel the physical basis of the same.

To understand the physical basis of the densely packed non-random arrangement of all CTs in a confined 3D nucleus, traditional microscopy based methods are complemented with {\it in silico} approaches using polymer-based models (representative models reviewed in \cite{Fudenberg_2012,Halverson_2011,Renom_Mirny_2011,Vasquez_Bloom_2014}). Interestingly, microscopy based studies of normal cells have confirmed that CTs do not spontaneously form knots, entangled within or across each other, and therefore it was suggested that their topology dictates such a segregated spatial arrangement \cite{Rosa_2008, Vettorel_2009}. Subsequently, it has been demonstrated that topological constraints govern discrete territorial clustering, because polymers without excluded volume and negligible intermingling have also revealed a radial-like clustering \cite{Blackstone_2011,Dorier_2009}. Additionally, it has been suggested that individual chromosomal activity derived from their intrinsic parameters (total number of genes, chromosome length and average gene density) govern radial arrangement \cite{Ganai_2014}. Interestingly, polymer-based models have used fixed parameters representing coding gene densities and / or chromosomal lengths \cite{Rosa_2008,Ganai_2014,Kreth_2004,Munkel_1998,Munkel_1999}, implying that the protein-coding genome may also constrain spatial CT arrangement. However, we lack a mathematical model that can address the couplings among intrinsic parameters, which pertain to individual chromosomes, to reveal a systems-level crosstalk, which is the focus of the current study. Additionally, the role of this systems-level crosstalk in the context of self-organized spatial arrangement of CTs is revealed.
\end{section}
\begin{section}{Results and Discussion}
Eukaryotic genomes encode the necessary blueprint for critical functions such as DNA replication, damage repair, transcription and RNA processing. Since the first completely sequenced human genome was made available for scientific research \cite{Lander_2001_hgenome_Nature,Venter_2001_hgenome_Science}, the traditional schematic of our genome is represented as a linear array of sequenced genetic information, with twenty-four CTs in sequential order (chr1, chr2, .. , chr21, chr22, chrX, and chrY). Here, we reveal that eukaryotic genomic information can also be represented as a matrix in abstract vector space, an abstract construct that can mathematically describe extrinsic genomic couplings among the constituent CTs. Our model reveals an overall unique non-random hierarchical arrangement among CTs, which arises along with permissible degeneracy (ambiguity due to permutations as in statistical physics). Using systems biology, we identify the physical principles that help constrain these CT arrangements in a normal human nucleus with forty-six chromosomes ($46$, XY).

{\bf \it{Unification within human genome - a systems biology approach.}}
For brevity, we denote human genome's N different CTs (chr1, chr2, .. , chr21, chr22, chrX and chrY) as $C_j$ (where $1 \leq j \leq N=24$). The gene density of $j^{th}$ CT ($C_j$) is represented by $d_j$ by the scalar equation $n_{j}=d_{j} L_{j}$, where $n_j$ is the total number of genes (that includes protein-coding and non-coding genes) and $L_j$ is the length of $C_j$. As $n_j$ is specified independent of $L_k$ even if $C_k$ is the nearest neighbor of $C_j$, $n_j$ is directly proportional to $L_j$, and constant of proportionality is the coupling constant $d_j$. Therefore, equations involving all intrinsic chromosomal couplings may be represented as $n_j= \sum\limits_{\substack{k=1}}^{k=N} \delta_{jk}d_{k}L_{j}$, where $\delta_{jk}$ is the Kronecker delta function ($=1$ for $j=k$, and $=0$ for $j\neq k$). We propose $\hat{\bD}$ as as a diagonal $N \times N$ matrix with intrinsic gene density ($d_j$) as the diagonal elements (real eigenvalues of $\hat{\bD}$). It must be noted that intrinsic parameters are experimentally obtained from cytogenetic as well as sequencing efforts under laboratory conditions. Therefore, these parameters may not accurately represent the systems biology of chromosomal milieu that nuclei manifest {\it in vivo}. For brevity, we denote the set of basis vectors for the {\it in vitro} or diagonal representation as: $\{\ket{n_{1},L_{1}}, \ket{n_{2},L_{2}}, \cdots , \ket{n_{N},L_{N}} \}$ labeled by the intrinsic parameters: total gene count per chromosome ($n$) and its length ($L$) in megabase pair (Mbp) and a subscript $j$ for referencing $CTj$.  A systems-level formalism in a $N$ dimensional vector space using a $N \times N$ {\it in vitro} coupling matrix is presented as a matrix equation:
\begin{equation}
\label{eq:SB_diagD}
\ket{n}= \hat{\bD} \ket{L},
\text{where,}~ \hat{\bD}= 
\begin{pmatrix}
d_{1}  & 0      & \cdots & 0 \\
0      & d_{2}  & \cdots & 0 \\
\vdots & \vdots & \ddots & \vdots \\
0      & 0      & \cdots & d_{N} \\
\end{pmatrix}
\end{equation}
such that the intrinsic gene density $d_j = \matrixelement{n_{j},L_{j}}{\hat{\bD}}{n_{j},L_{j}}$ (Mbp$^{-1}$ units). Here, $\ket{n}$ and $\ket{L}$ are $N \times 1$ column vectors whose components represent the {\it in vitro} intrinsic parameters of number of coding genes and chromosome length. As the {\it in vitro} gene density space does not reveal inter-chromosomal couplings, the hierarchical nature of extrinsic gene density remains implicit or ``hidden''. Next, we formulate a mathematical scheme to describe extrinsic chromosomal parameters in an {\it in vivo} context.

We describe biological crosstalk among all the CTs in the human nucleus using a gene density matrix that represents extrinsic couplings among all chromosomes. Systems-level coupling among them may be represented in a $N$ dimensional {\it in vivo} vector space defined by a $N \times N$ Hermitian matrix $\hat{K}$. This space is defined by a set of $N$ basis vectors $\{\ket{n_{1}^{'},L_{1}^{'}}, \ket{n_{2}^{'},L_{2}^{'}}, \cdots , \ket{n_{N}^{'},L_{N}^{'}} \}$ labeled by extrinsic systems-level parameters: effective gene count ($n^{'}$) and effective length ($L^{'}$) in Mbp, along with subscripts that denote labels from column and row vectors. Next, we represent an abstract effective gene count and effective chromosomal length as vectors denoted by $\ket {\mathbb{N}}$ and $\ket{\Lambda}$  respectively, and analogous to the {\it in vitro} model (Equation \ref{eq:SB_diagD}), we posit:
\begin{equation}
\begin{split}
\label{eq:SB_PCGC}
\noindent{
\ket{\mathbb{N}}= \hat{K} \ket{\Lambda}}, ~~~~~~~~~~~~~~~~~~~~~~~~~~~~~~~~~~~~~~~~~~~~~~~\\
\text {or,~~}
\begin{pmatrix} \hat{\mathbb{N}} \ket{n_{1}^{'}L_{1}^{'}}  \\
  \cdots \\
  \hat{\mathbb{N}} \ket{n_{N}^{'}L_{N}^{'}}  \\
\end{pmatrix} = ~~~~~~~~~~~~~~~~~~~~~~~~~~~~~~~~~~~~~~~~~~~~~~~~~~~~~~~~~~~~~~~~~~ \\
\begin{pmatrix} 
\matrixelement{n_{1}^{'}L_{1}^{'}}{\hat{K}}{n_{1}^{'}L_{1}^{'}} &\cdots  &  \matrixelement{n_{1}^{'}L_{1}^{'}}{\hat{K}}{n_{N}^{'}L_{N}^{'}}  \\
\cdots                   & \cdots & \cdots \\ 
\matrixelement{n_{N}^{'}L_{N}^{'}}{\hat{K}}{n_{1}^{'}L_{1}^{'}} &\cdots  &  \matrixelement{n_{N}^{'}L_{N}^{'}}{\hat{K}}{n_{N}^{'}L_{N}^{'}}  \\
\end{pmatrix}
\begin{pmatrix} \hat{\Lambda} \ket{n_{1}^{'}L_{1}^{'}}  \\
  \cdots \\ 
  \hat{\Lambda} \ket{n_{N}^{'}L_{N}^{'}}  \\  \end{pmatrix} 
~~~~~~~~~~~~~
\end{split}
\end{equation}
We hypothesize that the systems-level coupling among CTs is commutative and associative, such that it may be represented by a matrix $\hat{K}$, which is  real, symmetric Hermitian matrix. For a minimalistic approach, we hypothesize that the effective gene density is due to nearest-neighbor inter-CT coupling of $C_j$ with $C_k$, and thus we approximate the original generic coupling matrix $\hat{K}$ in Equation \ref{eq:SB_PCGC} with $\hat{\bSigma}$:
\begin{equation}
\begin{split}
\label{eq:SB_pairCT_aprox}
\hat{K} & \approx \hat{\bSigma} = \\
& \begin{pmatrix} 
\matrixelement{n_{1}^{'}L_{1}^{'}}{\hat{\bSigma}}{n_{1}^{'}L_{1}^{'}} &\cdots  &  \matrixelement{n_{1}^{'}L_{1}^{'}}{\hat{\bSigma}}{n_{N}^{'}L_{N}^{'}}  \\
\cdots                   & \cdots & \cdots \\ 
\matrixelement{n_{N}^{'}L_{N}^{'}}{\hat{\bSigma}}{n_{1}^{'}L_{1}^{'}} &\cdots  &  \matrixelement{n_{N}^{'}L_{N}^{'}}{\hat{\bSigma}}{n_{N}^{'}L_{N}^{'}}  \\
\end{pmatrix}
\end{split}
\end{equation}
Next, we propose that extrinsic gene density of the system may be derived using its intrinsic chromosomal parameters: number of genes ($n_j$), chromosomal length ($L_j$), gene density ($d_j$), and $N$ scalar equations $n_j= \sum\limits_{\substack{k=1}}^{k=N} \delta_{jk}d_{k}L_{j}$ with $d_{j} > 0$, $L_{j} > 0$, and where $\delta_{jk}$ is the Kronecker delta function. To explicitly reveal the inter-CT ``mixing'' (which we refer to as biological crosstalk), we represent $\hat{\bD}$ in the {\it in vivo} vector space, which is defined by {\it in vivo} basis vectors. This deconstruction is obtained, using the Spectral Theorem, as a symmetric $N \times N$ matrix $\hat{\bSigma}$ (from Equation \ref{eq:SB_pairCT_aprox}) such that
\begin{equation}
\label{eq:SpecTheorem}
\hat{\bSigma}=\hat{\bGamma}^{\mathsmaller T}\hat{\bD} \hat{\bGamma}.
\end{equation}
Here, the $j$ row and $k$ column element of $\hat{\bSigma}$  is denoted as $\sigma_{jk}^{'}$, in Mbp$^{-1}$ units, (where $\matrixelement{n_{j}^{'},L_{j}^{'}}{\hat{\bSigma}}{n_{k}^{'},L_{k}^{'}}$) and $\hat{\bGamma}$ is a $N \times N$ orthogonal matrix ($\hat{\bGamma}^{\mathsmaller T} \hat{\bGamma}= \hat{\aUnit}_{N \times N}$ =unitary matrix). For an ordered CT pair of $C_j$ and $C_k$, denoted $C_{jk}^{'}$, we derive $\sigma_{jk}^{'}$, the $(j,k)$ element of the effective gene density matrix ($\hat{\bSigma}$) as (Materials and Methods):
\begin{equation}
\label{eq:eff_sigma_p1}
\sigma_{jk}= \frac{d_{j}d_{k}}{(d_j+d_k)} \left[ { 1+ \frac{L_{j}d_{k}+L_{k}d_{j}}{L_{j}d_{j}+L_{k}d_{k}} } \right].
\end{equation}
This effective gene density parameter $\sigma_{jk}^{'}$ represents a minimal extrinsic coupling to an effective chromosomal length ($L_{jk}^{'}$) for $C_{jk}^{'}$, and is represented in a histogram and heatmap in Fig. \ref{Fig_PCGC_v1}A and 1B respectively. The histogram for effective gene density is asymmetric, skewed toward higher values. The highest effective gene density for $C_{chr19,chr19}^{'}$ is statistically significant from the histogram mean (difference from mean is $4.7$ times RMSD of histogram). Effective gene density heat map reveals that for a given CT, there may be variability in its extrinsic gene density, that is contingent on neighborhood of CTs. For example, effective gene density couplings for $C_{chr19,chrY}^{'}$ are much lower (less than fifty percent) the effective gene density couplings for $C_{chr19,chr17}^{'}$.

To represent a systems-level paired chromosome construct, we also computed an effective number of genes for $\CTjk$, which we termed ``paired chromosome's gene count'' (PCGC) $(n_{jk}^{'}$, Materials and Methods, Equation \ref{eq:PCGC_effL_effsigma}) and also defined a genome-specific normalized PCGC $(\pi_{jk}$, Materials and Methods, Equation \ref{eq:PCGC_main_norm}). The effective number of gene count for $\CTjk$ is revealed by the coupling of effective gene density with an effective length (Equation \ref{eq:effL_hm}). Using PCGC, we wanted to determine if neighborhood CT effects could be used to delineate spatial CT arrangements. Hence, we sought to identify patterns of effective gene density among CTs, the spatial positions of which were already known from high volume microscopy study performed at our laboratory using dermal fibroblast nuclei \cite{Mehta_2013}. Therefore, we computed $\pi_{jk}^{'}$ versus $L_{jk}^{'}$ (effective length) for cases when $C_j$ and $C_k$ were both exclusive to (i) the nuclear interior, (ii) the periphery, and (iii) spatially intermediate to (i) and (ii). Such a scatter-plot of $\pi_{jk}^{'}$ versus $L_{jk}^{'}$ (Fig. \ref{Fig_PCGC_v1}C), whose composition is cell-type specific, reveals segregation among paired CTs that are in the inner core versus the periphery of the nucleus, suggesting a hierarchy in PCGG versus effective length for $\CTjk$. The scatter-plot retrieved for fibroblast also reveals intermediate category, which depends on whether $\CTj$ and $\CTk$ is interior or peripheral. It is important to note that the composition of interior versus periphery undergoes a subtle change in lymphocyte (as shown in Table \ref{tab1}), leading to commensurate changes in the intermediate category (data not shown). Interestingly, for $\CTjk$ that do not overlap with instances where both CTs are confined to the interior / intermediate / periphery of the nucleus are also represented in the scatter-plot. A CT from any one of the above three spatial zones coupled with another CT from a different zone constitutes the non-overlapping category (Fig. \ref{Fig_PCGC_v1}C). Therefore, we hypothesize that a systematic hierarchy and degeneracy in spatial CT arrangement is obtained when inter-chromosomal coupling via effective gene density is computed. We systematically analyze this feature further.

{\it Effective gene density corroborates spatial CT positioning in human interphase nucleus.}
As normalized PCGC $(\pi_{jk})$ and effective chromosomal length $(L_{jk}^{'})$ terms are coupled via the effective gene density $\sigma_{jk}^{'}$ (Equation \ref{eq:PCGC_main_norm}, Materials in Methods), we investigate its hierarchy in the context of CT arrangements in human interphase nuclei. We used complete-linkage algorithm implemented in R software-package \cite{R_prog} to reveal the hierarchical clustering within effective gene density space. Our analysis of the human genome led to a segregation of effective gene density from CTs into two primary clusters Groups A and B that are represented in Fig. \ref{Fig_PCGC_v1}B, and in Fig. \ref{Fig_hclust}. We denote Group A with chr11, chr16, chr17, chr19, chr20, chr22 and Group B with the remaining 18 CTs (Fig. \ref{Fig_hclust}). If we consider a hypothetical pruning of chr19 (from Group A), then the hierarchy of Group A changes with respect to Group B (the common branch node for chr19 with other Group A members is highest). The primary hierarchy of chr19 obtained using complete-linkage algorithm (Fig. \ref{Fig_hclust} and Fig. S1A), was also confirmed by centroid-linkage (Fig. S1B), average-linkage (Fig. S1C), single-linkage (Fig. S1D), and Mcquity (Fig S1E) algorithms implemented in R. Our results imply that primary hierarchy of chr19 in effective gene density space is necessary for the overall non-random and hierarchical spatial arrangement of other CTs in the human nucleus. Therefore, Group A was subdivided into Subgroup A$'$ (chr19) and Subgroup A$''$ (chr11, chr16, chr17, chr20 and chr22) (Fig. \ref{Fig_hclust}). 

As chr19 signified a unique and primary hierarchy among all $C_{j,k}^{'}$, one could topologically consign it to be at the origin of an abstract effective gene density space, or consign it to the interior of the nuclear volume (chr19 is at the core interior in both fibroblast and lymphocyte nuclei \cite{Boyle_2001,Mehta_2013}). This one-to-one mapping between effective gene density space with the real spatial nuclear volume facilitated subsequent characterization of spatial arrangement of the other CTs with respect to chr19, or its immediate neighbors $C_{chr19,k}^{'}$, in a logical sequence as if it were a ``constellation'' of CTs (CT constellation from the interior toward the periphery). Thus, the theoretical formalism derived from this systems biology theory encompasses principles that give rise to a self-organized CT arrangement within the nucleus, substantiating a previously proposed hypothesis \cite{Misteli_2001_JCB_rev,Misteli_2009_PNAS_rev}. As our results revealed that the hierarchy of Subgroup A$''$ was juxtaposed intermediate to chr19, and Group B (Fig. \ref{Fig_hclust}), this implied that CTs of Subgroup A$''$ (chr11, chr16, chr17, chr20 and chr22) were constrained to be adjacent to Subgroup A$'$ (chr19) in effective gene density space in a non-random constellation. In addition, this scheme reveals degenerate representations of CTs 11 and chr16 in the neighborhood of other CTs (Fig. \ref{Fig_hclust}). All remaining eighteen CTs (Group B), which constituted a dominant fraction, were largely clustered off from Group A in the context of chr19, except for chrY. The dendrogram leaf representing chrY stands out juxtaposed between the six Group A CTs and seventeen Group B CTs (Fig. \ref{Fig_hclust}). Noticeably, on the heat map of Fig 1B, gradation of effective gene density for $C_{chrY,k}^{'}$ pairs are relatively lower and more uniform (lesser variability) compared to say for example $C_{chr19,k}^{'}$ where there is greater variability. (Note the ``corridor'' shown by dashed and dotted lines in Fig. \ref{Fig_PCGC_v1}B). Clearly, if chrY were pruned from the dendrogram, the hierarchy of Group A with respect to Group B would not alter. Thus chrY results support three important consequences of CT arrangements in human nuclei. Firstly, it reveals that the physical basis for CT arrangement in $46$XX versus $46$XY nuclei is identical. Secondly, contingent on near-neighbor CTs via effective gene density couplings, cell type, and the poised state of genes on CTs, the spatial position of chrY may be in the neighborhood of chr19 (Fig. \ref{Fig_hclust}A), or it may as well be entirely distant from it (Fig. \ref{Fig_hclust}B and \ref{Fig_hclust}C). Thirdly, and most importantly, the ``orphan'' nature of chrY (for example, with no homologous partner for crossing over, and its functional specialization for spermatogenesis) is also evident from our systems-level effective gene density hierarchy analyses.

{\it Extrinsic systems-level constraints greatly exceed intrinsic constraints.}
Our minimalistic $N$ dimensional vector model treats $\CTjk$ as unified entities instead of solitary ones. In Equation \ref{eq:eff_sigma_p1}, the derived effective gene density for $\CTjk$ has ``like'' terms $(L_{j}, d_{j})$ and ``unlike'' terms $(L_{j}d_{k})$, representing a systems-level coupling of CTs. From $\hat{\bSigma}$ up to a total of $\frac{1}{2}N(N+1)$ unique $\sigma_{jk}^{'}$ values from $\CTjk$ are obtained, each of which is described by two parameters: effective length and effective gene density. Therefore, up to $N(N+1)$ unique equations constrain the coding genome, from which only $2N$ are intrinsic. Hence, for the human nucleus, the maximum possible number of extrinsic systems-level constraints, via this minimal {\it in vivo} model, is nearly an order of magnitude larger (twelve times) than that from an {\it in vitro} model. Thus, this vector model reveals a biological crosstalk, implying a tighter regulation of chromosomal length with gene density for both homologous and non-homologous chromosome partners within the nucleus. In theory, if one were to extend this model to incorporate three near-neighbor CT interactions (in Equation \ref{eq:SB_pairCT_aprox}), additional genomic constraints will be revealed, implying an even tighter regulation among those extrinsic parameters.

{\it Previous empirical reports and a resolution of apparent anomalies.} Experiments have determined CTs with high gene density occupy the nuclear interior, but those with low gene density are toward the periphery \cite{Boyle_2001,Croft_1999,Mehta_2013}. It has also been reported that relatively small sized CTs are localized to the nuclear interior, but larger ones are toward the periphery \cite{Bolzer_2005,Cremer_2001,Sun_2000}. At an empirical level, both intrinsic parameters: gene density and chromosome size seem to predict CT arrangement (reviewed in \cite{Cremer_NatRev_2001,Foster_2005}), barring exceptions. Instances of inconsistent CT location for two widely studied cell types: fibroblasts and lymphocytes are presented in Table \ref{tab1}. Here we highlight some notable caveats related to CT arrangements: (1) spatial positions of chr21 and chrY, both small-size CTs with low gene density, are inconsistent in fibroblast and lymphocyte nuclei. Both CTs are largely at the nuclear interior in fibroblasts but also at the periphery in lymphocytes, (2) inconsistency in the spatial locations of seven other CTs (chr1, chr5, chr6, chr8, chr15, chr16 and chr20) in fibroblasts versus lymphocytes, (3) consistent (or weakly consistent) consensus locations obtained for chr10, chr11 and chr14 in these two cell types. Interestingly, unlike most CTs, these CTs do not seem to have a preferred spatial territory in the nuclei (interior versus intermediate versus periphery), (4) most importantly, in the realm of simplistic gene density or size-based segregation, the threshold value of high versus low intrinsic parameter that may lead to segregation of interior versus peripheral CTs is largely qualitative or at best empirical, and lacks mathematical rigor. Therefore, the physical basis of spatial CT arrangement in a densely packed milieu of human nucleus is highly coarse-grained and inadequate when described using intrinsic parameters.

{\it Rationalization of degenerate CT arrangement using PCGC.} We have demonstrated that effective gene density explicitly reveals hierarchy, degeneracy, and constrains spatial CT arrangement. The PCGC model delineated seventeen CTs from Group B (excluding chrY) as largely equidistant {\it en masse} from Group A, (primarily in the context of chr19) in effective gene density space. However, due to degenerate configurations, it is plausible that cohorts of CTs in Groups A and B will have an effective gene density-based altered spatial arrangement (again, primarily in the context of chr19), and still maintain effective gene density hierarchy. This hierarchical and degenerate representation enabled non-unique yet non-random chromosomal spatial arrangement, which has been experimentally discerned but not rationalized so far. As mentioned earlier, the position of chrY in the dendrogram facilitates its spatial locations in neighborhood of six CTs from Group A (largely gene rich CTs) as well as the other seventeen CTs from Group B (largely gene poor CTs) as represented in the two equivalent dendrogram representations Fig. \ref{Fig_hclust}A and \ref{Fig_hclust}B. Similarly, it can also be argued that near-neighbor effects will give rise to different CT representations for chr21 in effective gene density space, the outcome of which can give rise to contrasting spatial positions (interior versus periphery) in fibroblasts and lymphocytes (Fig \ref{Fig_hclust}).

As chr11 is consigned to a more spatial interior in certain CT ``constellation'' (Fig. \ref{Fig_hclust}A), we extend the rational for its degenerate spatial location by positioning it with chr1 and chr14 as intermediate to peripheral (Fig. \ref{Fig_hclust}B).  For example, chr11 in the vicinity of chr1 or chr14 (Fig. \ref{Fig_hclust}B) has lower effective gene density, as opposed to chr11 in the vicinity of chr16 or chr20 (Fig. \ref{Fig_hclust}C) as CTs couple differently among themselves in these two scenarios (a lower effective gene density neighborhood results if the gene density disparity among them is high and vice versa). Clearly spatial positioning of CTs can be very different, particularly when cell-specific instances and poised state of genes are considered, and therefore supporting self-organization. Moreover, permutations of CTs within degenerate effective gene density neighborhoods will have negligible consequence at a global scale in the nucleus. Hence, such CT permutations may be more feasible as opposed to those involving non-degenerate effective gene density neighborhoods. Additionally, the topology of CT pair is governed by neighborhood CT effects via the effective length (\ref{eq:effL_hm}), again reinforcing self-organization among CTs. Therefore, this unified theory of the genome, without any free parameter, is general enough for applicability to any annotated eukaryotic genome to predict its non-random hierarchical and degenerate spatial map of CT positions, in contrast to a simple chromosome length-based or gene density-based model.

Matrix methods have been extensively used in various scientific and engineering disciplines, including biological sciences, to reveal systems-level information that is usually implicit in data. For example, differential gene expression obtained from microarray data, defined in an abstract space specified by ``eigengenes'' and ``eigenarrays'' via vectors and matrices, revealed co-expressed over-active and under-active regulatory genes in genome-wide studies  \cite{Alter_Botstein_2000_PNAS}.

In summary, the derived effective gene density (obtained for the human genome) unified the intrinsic gene density among pairs of CTs, set constraints on their spatial arrangement and revealed degeneracies within their spatial arrangement. The anisotropy associated with effective gene density space, in the {\it in vivo} representation, provides the physical basis for a non-random self-organized spatial arrangement of CTs even in a crowded densely packed milieu of nuclear space. We corroborated and rationalized our findings with available experimental data for the human nucleus, to support our hypothesis that our genome constrains interphase spatial arrangements of the CTs in the nucleus (via effective gene density). We surmise that, within the constraints of hierarchy prescribed in the current model, the underlying biology posits finer mapping of CTs in cell-type specific instances. This unified theory, which requires no additional experimental input, may be extended to other eukaryotic species with annotated genomes to infer the physical principles that constrained self-organizing spatial CT arrangements as their genomes evolved.
\end{section}

\begin{section}{Materials and Methods}
{\it Intrinsic parameters.}
The intrinsic parameters of the human genome: total number of annotated genes per chromosome, and their respective lengths were obtained from National Center of Biotechnology Information (NCBI) Gene database \cite{NCBI_NAR_2015} and Mapviewer portal (release 107) for human genome \cite{NCBI_NAR_2015}.

{\it Extrinsic parameters and systems-level effective gene density.}
We hypothesize that for each nearest-neighbor CT pair $\CTjk$, the effective gene count $(n_{jk}^{'})$ or paired chromosome's gene count (PCGC) is coupled to an effective length $L_{jk}^{'}$ via an effective gene density $(\sigma_{jk}^{'})$ as:
\begin{equation}
\label{eq:PCGC_effL_effsigma}
n_{jk}^{'}= \sigma_{jk}^{'} L_{jk}^{'}.
\end{equation}
We define the dimensionless PCGC parameter $n_{jk}^{'}$ as the harmonic mean of annotated genes in $\CTjk$. The harmonic mean (as opposed to geometric mean or arithmetic mean) is the best representation for number of genes per unit length, as it gives lower weightage to very high values, when $n_j$ and $n_k$ are markedly dissimilar for given $\CTjk$.  Therefore:
\begin{equation}
\label{eq:PCGC_hm}
n_{jk}^{'}= \frac{2n_{j}n_{k}}{(n_j + n_k)}.
\end{equation}
and similarly, to best represent diverse lengths, we define $L_{jk}^{'}$ as the harmonic mean of $L_j$ and $L_k$:
\begin{equation}
\label{eq:effL_hm}
L_{jk}^{'}= \frac{2L_{j}L_{k}}{(L_j + L_k)}.
\end{equation}
On representing $d_{jk}^{'}$ as harmonic mean of intrinsic gene densities:
\begin{equation}
\label{eq:effdensity_hm}
d_{jk}^{'}= \frac{2d_{j}d_{k}}{(d_j + d_k)},
\end{equation}
and using Equations \ref{eq:PCGC_effL_effsigma} - \ref{eq:effdensity_hm}, we derive effective gene density as
\begin{equation}
\label{eq:eff_sigma_p2}
\sigma_{jk}= \frac{d_{jk}^{'}}{2} \left[ { 1+ \frac{L_{j}d_{k}+L_{k}d_{j}}{L_{j}d_{j}+L_{k}d_{k}} } \right] = \sigma_{kj}.
\end{equation}
This equation, together with Equation \ref{eq:effdensity_hm}, gives us Equation \ref{eq:eff_sigma_p1}. To normalize $n_{jk}^{'}$ in Equation \ref{eq:PCGC_effL_effsigma}, we used $T_{PCGC}~\big(= \sum\limits_{\substack{j=1}}^{N} \sum\limits_{\substack{k= j}}^{N} p_{jk}^{'} \big)$, and $M~\big( =\frac{1}{2}N(N+1) \big)$. Therefore, if expressed as a percentage, normalized Equation \ref{eq:PCGC_effL_effsigma} reads as:
\begin{equation}
\label{eq:PCGC_main_norm}
\pi_{jk}= \frac{n_{jk}^{'}}{MT_{PCGC}} 100\% = \frac{\sigma_{jk}L_{jk}^{'}}{MT_{PCGC}} 100\%.
\end{equation}
The scatter plot of $\pi_{jk}$ (normalized PCGC) versus effective length suggested hierarchy and degeneracy of effective gene density (Fig. \ref{Fig_PCGC_v1}C).
\end{section}

{\appendix
\section{APPENDIX: Abbreviations, symbols and mathematical representations in vector space}
\noindent
CT = chromosome territory, \\
Mbp = megabase pair, \\
PCGC = paired chromosome's gene count, \\
RMSD = root mean square deviation, \\
In this report, $\hat{M}$ denotes a matrix, $\ket{C}$ denotes a column vector, and $\bra{R}$
denotes a row vector, such that $\hat{M}\ket{C}$, $\bra{C}\hat{M}$, and $\innerproduct{R}{C}$
denote inner products and  $\dyad{C}{R}$ denotes an outer product. The labels $C$ and $R$ are parameters (intrinsic / extrinsic parameters) representing the vectors.
}


{\section{Figure Legends}{
\noindent
FIG. \ref{Fig_PCGC_v1}. Effective gene density of paired chromosome constructs from the human genome. Histogram of effective gene density from unique $\CTjk$ chromosome pairs are shown in panel A. Panel B shows a heatmap of effective gene density representing $\CTjk$ chromosome pairs, and are depicted by labels $j$ and $k$ on the $y$- and $x$-axes respectively. Values of effective gene density from $\CTjk$ and $\CTkj$ are identical across the dotted diagonal line and the horizontal and vertical dashed lines segregate CT pairs where both $C_j$ and $C_k$ are from Group A, versus those from Group B, or their admixture. The color key for effective gene density value is consistent in panels A and B. Panel C represents the graph of normalized effective number of genes for $\CTjk$ $(\pi_{jk} )$ versus effective length $(\pi_{jk} )$ for all $\CTjk$ is shown as shaded circle $\circ$. The coordinates representing $\CTjk$ pairs where both CTs are from nuclear interior $(\delta )$, periphery $(+)$, and spatially intermediate region $(\star )$ are superposed over those representing all pairs $\circ$. The $\circ$ that do not overlap with $(\delta )$, $(+)$, or $(\star )$ represent those pairs where CTs belong to different spatial categories.

FIG. \ref{Fig_hclust}. Hierarchical clustering of effective gene density from paired chromosome constructs are shown as equivalent dendrograms. Interior and peripheral chromosomes are denoted as Group A and B respectively. Although chr21 and chrY are Group B CTs and relatively gene poor, the model also supported them as interior chromosomes due to neighborhood CT effects (panel A). A hierarchical and degenerate representation, via PCGC model, enabled the rationalization of chr21 and chrY as interior CTs (panel A) such as in fibroblasts versus peripheral CTs such as in lymphocytes (panel B) revealed from these equivalent dendrograms. Similarly, chr11, chr1 and chr14 were rationalized as intermediate to peripheral CTs (panel B versus panel C).
}}%

{\section{Supporting Information caption}{
\noindent
FIG. S1. Hierarchical clustering from five algorithms, for effective gene density among paired chromosome constructs in the human genome. Dendrograms reveal the primary hierarchy of chr19, as obtained for CT pairs using the systems-level paired chromosome's gene count (PCGC) model, using five algorithms: complete-linkage (panel A), centroid-linkage (panel B), average-linkage (panel C), single-linkage (panel D) and Mcquity method (panel E) implemented in R language and environment for statistical computing.
}}




%




\begin{acknowledgments}
This work was supported by funding from TIFR (Department of Atomic Energy, India) to SNF, BJR (Grant No. 12P-0123) and Sir JC Bose Award Fellowship from Department of Science and Technology, India to BJR (Grant No. 10X-217). We thank Professor G Ravindrakumar (TIFR) and Dr. Ravi Venkatramani (TIFR) for critical reading and suggestions to the manuscript.
\end{acknowledgments}



%

\clearpage
\begin{turnpage}
\begin{table}[t] 
\caption{\label{tab1} {\bf Location of CTs in human nuclei.}}
\begin{ruledtabular}
\begin{tabular}{l l l l}
{\bf chr ID} & {\bf Fibroblasts}$^{\dag}$ & {\bf Lymphocytes}$^{\P}$ & {\bf Summary} \\ 
\hline
chr1   & Intermediate - Periphery & Interior - Intermediate & \textbf{\textit{Inconsistent}} \\
chr2   & Periphery & Periphery    & Consistent \\
chr3   & Periphery & Periphery    & Consistent \\
chr4   & Periphery & Periphery    & Consistent \\
chr5   & Periphery & Intermediate & \textbf{\textit{Inconsistent}} \\
chr6   & Periphery & Interior - Intermediate - Periphery & \textbf{\textit{Inconsistent}} \\
chr7   & Periphery & Periphery    & Consistent \\
chr8   & Interior - Intermediate - Periphery & Periphery & \textbf{\textit{Inconsistent}} \\
chr9   & Periphery & Periphery    & Consistent \\
chr10  & Interior - Intermediate - Periphery & Intermediate & {\it Weakly Consistent} \\
chr11  & Periphery & Intermediate - Periphery & {\it Weakly Consistent} \\
chr12  & Periphery & Periphery    & Consistent \\
chr13  & Periphery & Periphery    & Consistent \\
chr14  & Interior - Intermediate - Periphery  & Interior - Intermediate - Periphery  & Consistent \\
chr15  & Periphery & Intermediate & \textbf{\textit{Inconsistent}} \\
chr16  & Intermediate - Periphery & Interior - Intermediate  & \textbf{\textit{Inconsistent}} \\
chr17  & Interior  & Interior     & Consistent \\
chr18  & Periphery & Periphery    & Consistent \\
chr19  & Interior  & Interior     & Consistent \\
chr20  & Interior - Intermediate  & Periphery & \textbf{\textit{Inconsistent}} \\
chr21  & Interior  & Interior - Intermediate - Periphery & \textbf{\textit{Inconsistent}} \\
chr22  & Interior  & Interior     & Consistent \\
chrX   & Periphery & Periphery    & Consistent \\
chrY   & Interior  & Periphery    & \textbf{\textit{Inconsistent}} \\
\end{tabular}
\end{ruledtabular}
{A summary of CT positions obtained from Fig. 3 in $^{\P}$Boyle, {\it et. al.} 2001 \cite{Boyle_2001}  and Additional file 1 $^{\dag}$Mehta, {\it et. al.} 2013 \cite{Mehta_2013} .}
\end{table}
\end{turnpage}



%
\begin{figure}
\begin{center}
\includegraphics[scale=0.70]{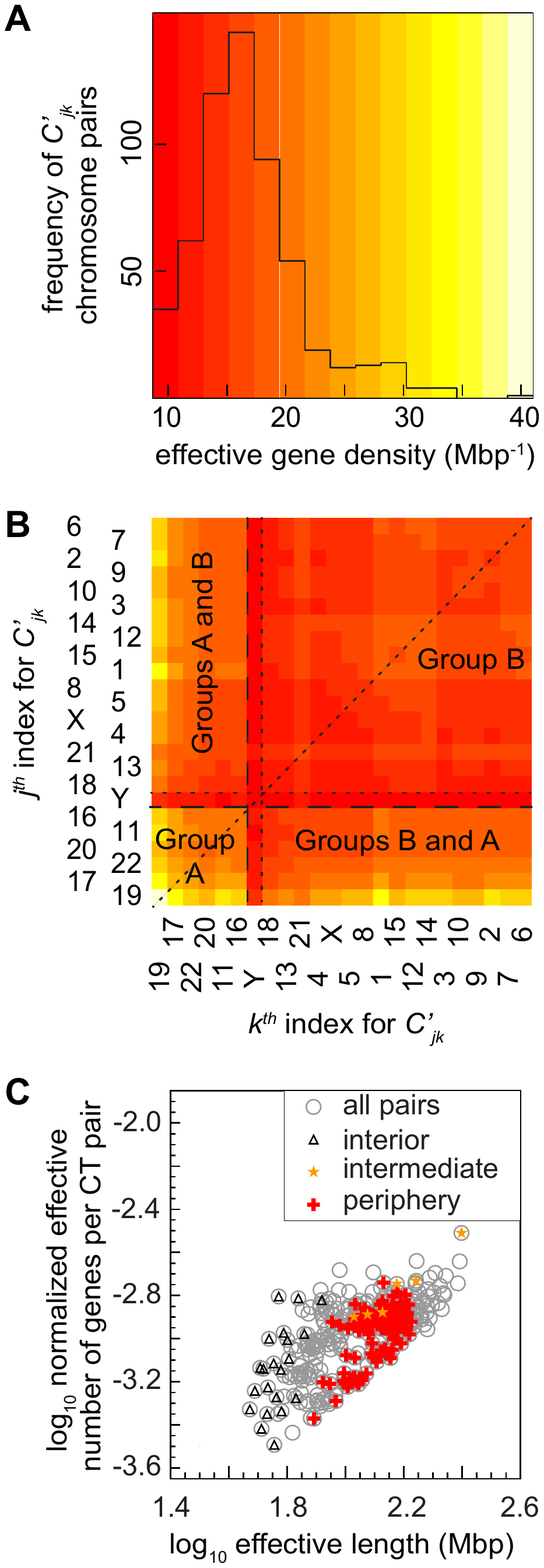} \\
\end{center}
\caption{\label{Fig_PCGC_v1} Effective gene density of paired chromosome constructs from the human genome.}
\end{figure}

\begin{figure}
\begin{center}
\includegraphics[scale=0.67]{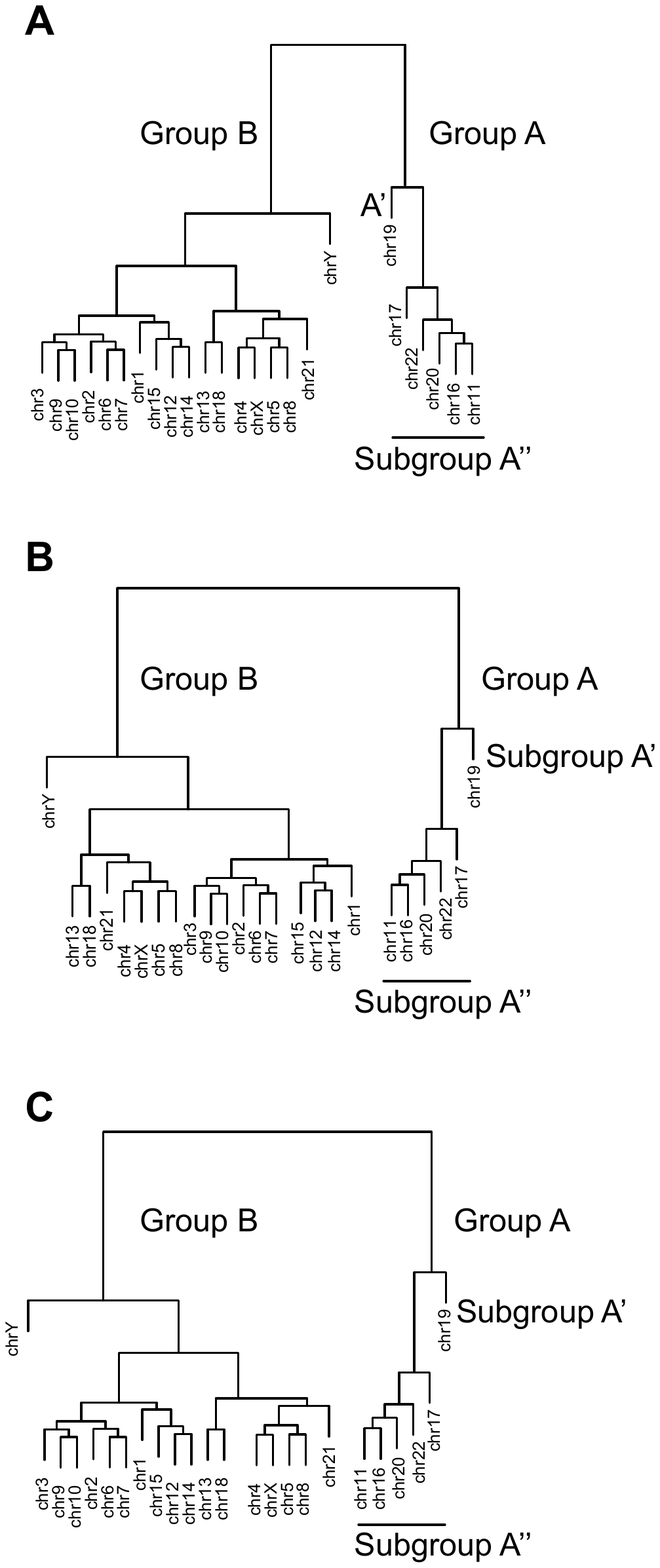} \\
\end{center}
\caption{\label{Fig_hclust} Hierarchical clustering of effective gene density from paired chromosome constructs are shown as equivalent dendrograms.}
\end{figure}

\begin{figure}
\begin{center}
\includegraphics[scale=0.67]{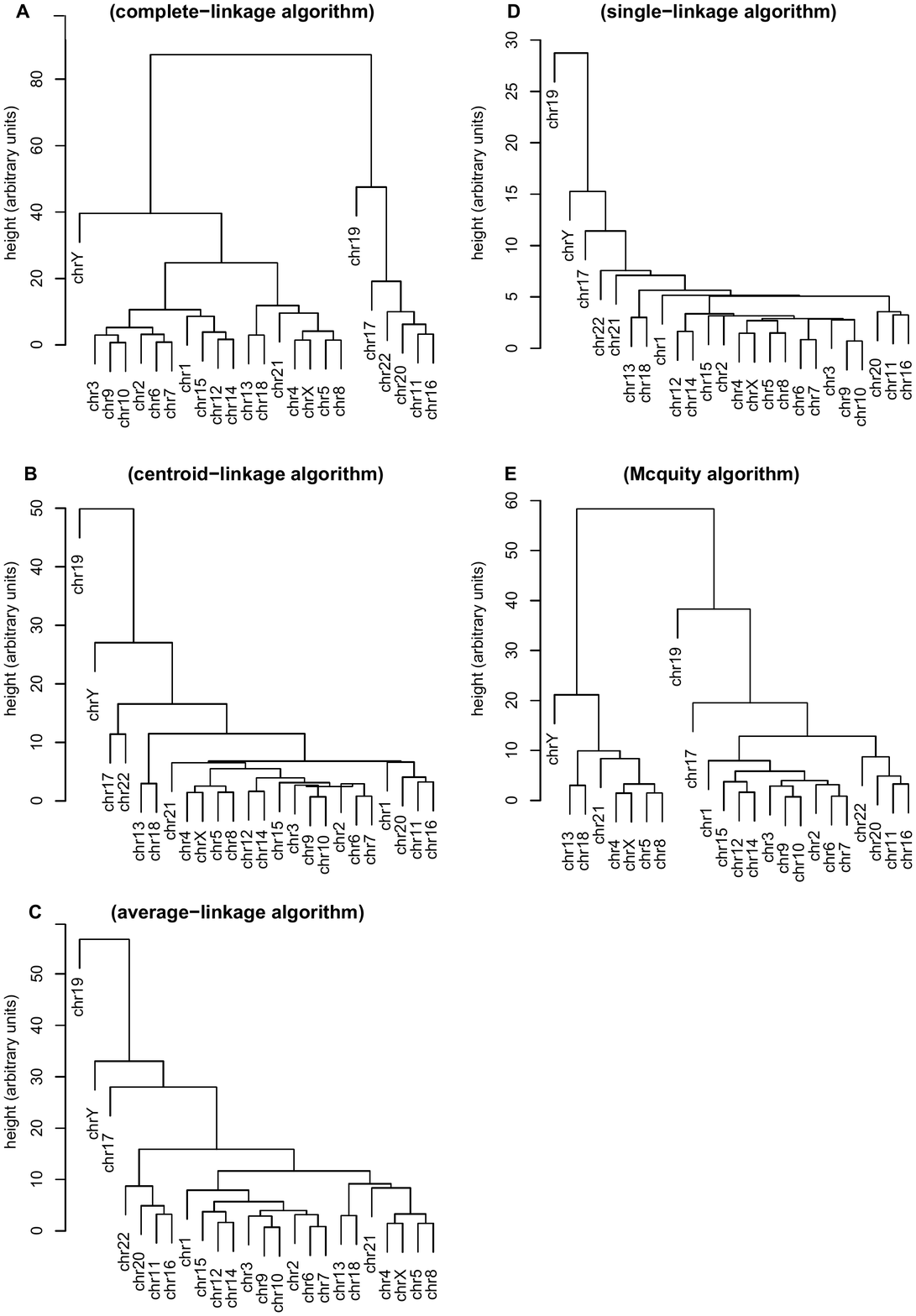} \\
\end{center}
\caption*{\label{Fig_hclust_models} FIG. S1. Hierarchical clustering from five algorithms, for effective gene density among paired chromosome constructs in the human genome.}
\end{figure}

\end{document}